\documentclass[aps,prb,twocolumn,amsmath,amssymb,floatfix,superscriptaddress]{revtex4-2}

\usepackage{graphicx,xcolor,multirow,bm,hyperref}
\usepackage{float}

\usepackage[ignoreunlbld,norefs,nocites]{refcheck}

\begin{document}

\title{Hot Electron-Driven Structural Expansion and Magnetic Collapse in Bilayer FeSe}
\author{Sam Azadi}
\email{sam.azadi@manchester.ac.uk}
\affiliation{Department of Physics and Astronomy, University of Manchester, Oxford Road, Manchester M13 9PL, United Kingdom}
\author{A.\ Principi}
\affiliation{Department of Physics and Astronomy, University of Manchester, Oxford Road, Manchester M13 9PL, United Kingdom}
\author{M.\ S.\ Bahramy}
\affiliation{Department of Physics and Astronomy, University of Manchester, Oxford Road, Manchester M13 9PL, United Kingdom}
\date{\today}
\begin{abstract}
Quantum phenomena emerging from the interaction of light and matter in low-dimensional systems hold great potential for future quantum technologies. Here, using first-principles calculations incorporating non-local van der Waals interactions and Hubbard corrections, we report simultaneous structural expansion and magnetic collapse in bilayer FeSe induced by photoexcited hot electrons. Our calculations reveal that, while bulk FeSe is paramagnetic, as observed experimentally, double-layer FeSe exhibits robust {\it staggered} antiferromagnetic order at low temperatures with a net site magnetization of $\sim 2.75~\mu_B$/Fe. However, increasing the density of photoexcited electrons systematically enhances the internal electronic entropy, leading to a complete collapse of antiferromagnetic order accompanied by an abrupt expansion of the interlayer separation. Our findings suggest the structural and magnetic properties of FeSe thin films can be finely tuned via ultrafast laser excitation, offering a pathway to control quantum phases in iron-based compounds through electronic temperature. 
\end{abstract}

\maketitle

Magnetic two-dimensional (2D) systems provide a versatile platform for exploring quantum phenomena and exotic phases of matter, with their reduced dimensionality significantly amplifying quantum effects \cite{HuaWang2022,Gibertini2019,Jenkins2022,Gong2017}. These materials exhibit strong magnetic anisotropy, which can be tuned through various external means, such as electric fields, strain, photoexcitation, and chemical doping \cite{Dagotto2013, Dai2015}. Moreover, their electronic and magnetic properties can be manipulated by controlling their thickness \cite{Huang2017}. Importantly, combining these systems with other 2D materials, such as graphene, creates heterostructures with novel functionalities, thereby opening new avenues for multifunctional devices \cite{Castro2009,GWang2018,Huang2020}.

The electronic properties of magnetic-layered systems are significantly influenced by interlayer separation, which is governed by long-range van der Waals (vdW) interactions \cite{Ding2021}. Controlling this separation, for example, through isothermal compression, allows the tuning of various emergent features, such as the coupling between nematicity and magnetism \cite{Kothapalli2016}. Additionally, strong local Coulomb interactions play a critical role in determining magnetic order in low dimensions. The coexistence of these competing orders, namely weak vdW forces and strong localized interactions in a 2D electronic system presents unique opportunities to explore exotic quantum phenomena, with implications for both fundamental science and technological applications.

Short-pulse light sources, such as femtosecond optical lasers, can generate ultrafast, non-equilibrium conditions in solids, making them powerful tools for exploring photo-induced phase transitions in strongly correlated systems. By rapidly driving electrons to high temperatures, these lasers can trigger nonthermal excitations that alter the low-energy density of states (DOS) and vibrational spectra \cite{Recoules, Rousse, Sciaini}. Such alterations can induce structural transitions below picosecond timescales, occurring well before significant energy transfer between electrons and ions \cite{Collet,Linderberg}. For instance, ultrafast electron crystallography has demonstrated that femtosecond laser pulses can induce structural changes in graphite, with first-principles calculations suggesting that the resulting non-thermal heating leads to interlayer expansion and compressive Coulomb stress from photo-induced charge separation \cite{Raman2008}.
\begin{figure}
    \centering
    \includegraphics[clip,width=1.0\linewidth]{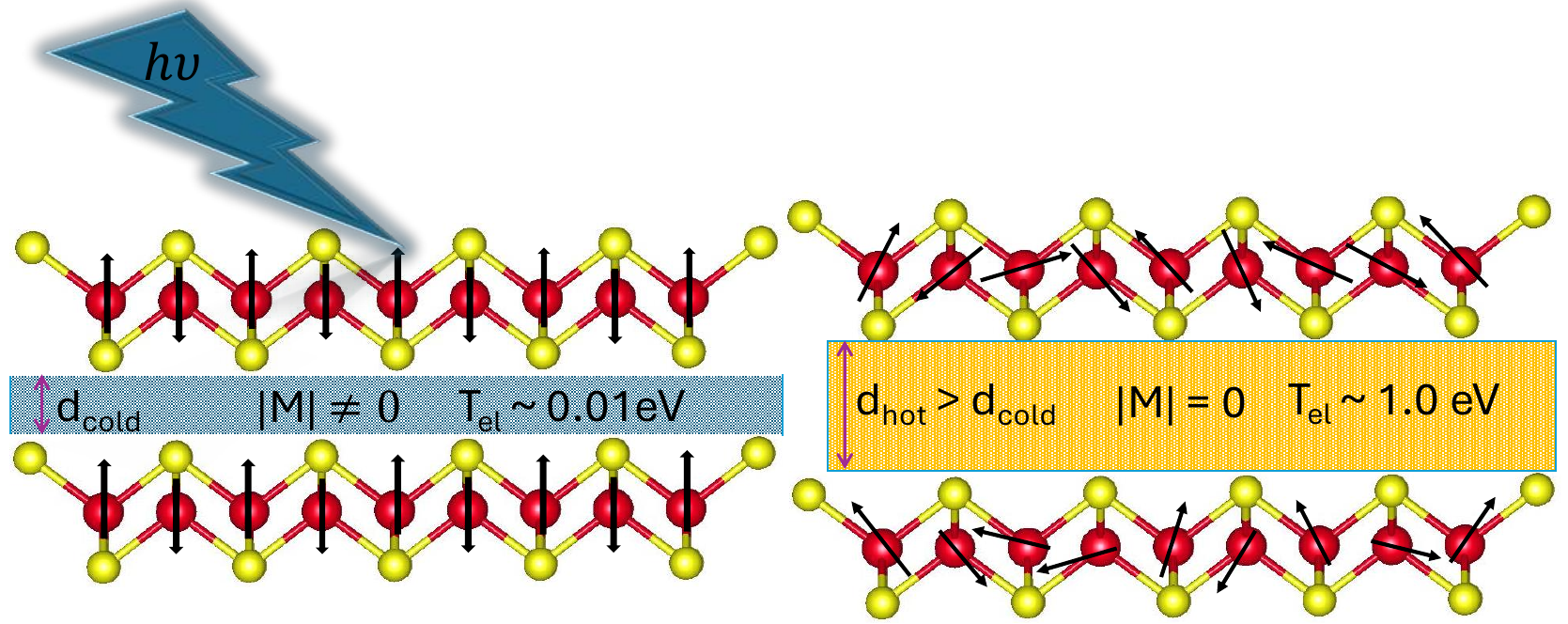}
    \caption{Double layer FeSe shows metallic antiferromagnetic properties at low temperature. Photo excited electrons causes an antiferromagnetic to paramagnetic phase transition and simultaneously the interlayer separation increases due to hot electron thermal pressure. Iron and selenium atoms are shown by red and yellow spheres, respectively.}
    \label{fig:fig1}
\end{figure}

In this work, we show that hot electrons generated by ultrafast photoexcitation can induce novel electronic, structural, and magnetic phase transitions in bilayer iron selenide (FeSe). We focus on FeSe, the prototypical member of iron chalcogenides, which has garnered significant theoretical and experimental attention \cite{MMa2017} as key building blocks for unconventional high-temperature superconductors based on transition metals \cite{Dai2015,McQueen2009,Alloul2009,MYi2013,ZYin2011}. FeSe exhibits intriguing features that suggest a complex interplay between nematicity, magnetism, and superconductivity. The bulk FeSe undergoes a structural phase transformation from tetragonal to orthorhombic at approximately $\sim90$ K \cite{Hsu2008,McQueen2009}, without displaying any magnetic order at ambient pressure. Its critical superconducting transition temperature is $\sim 8$ K \cite{Hsu2008}, which can increase to $\sim37$ K under a hydrostatic pressure of $\sim6$ GPa \cite{Medvedev2009,Okabe2010}. Notably, pressure also introduces magnetic ordering at approximately 1 GPa while simultaneously reducing the structural phase transition temperature \cite{Bendele2012, Bendele2010}.

\begin{figure*}[t]
    \centering
    \begin{tabular}{ccc}
        \includegraphics[scale=0.37]{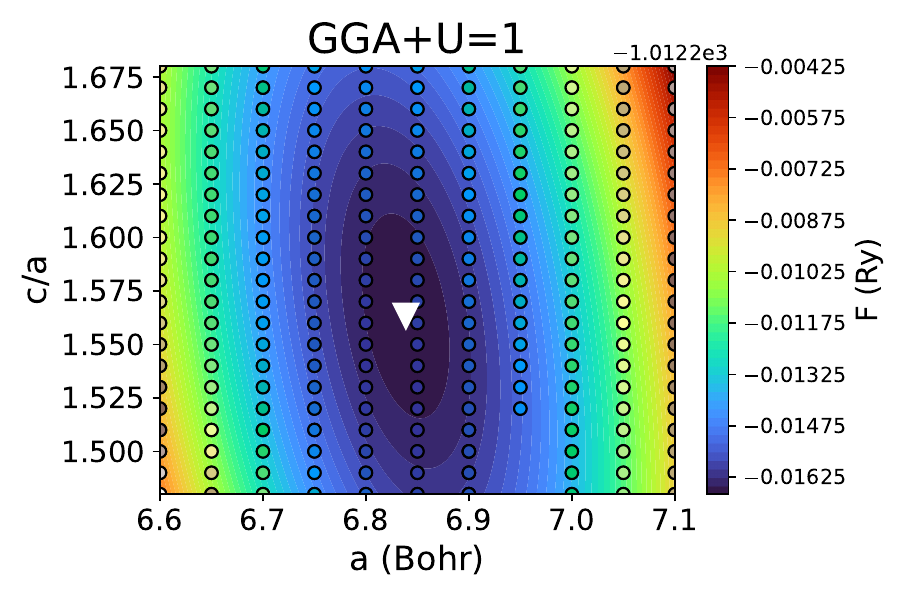}  & 
         \includegraphics[scale=0.37]{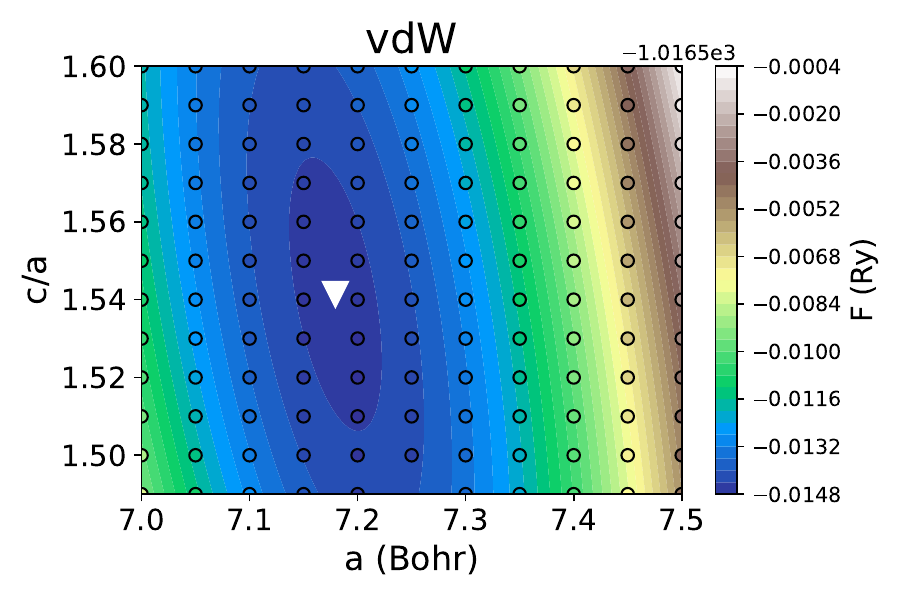} &  
         \includegraphics[scale=0.37]{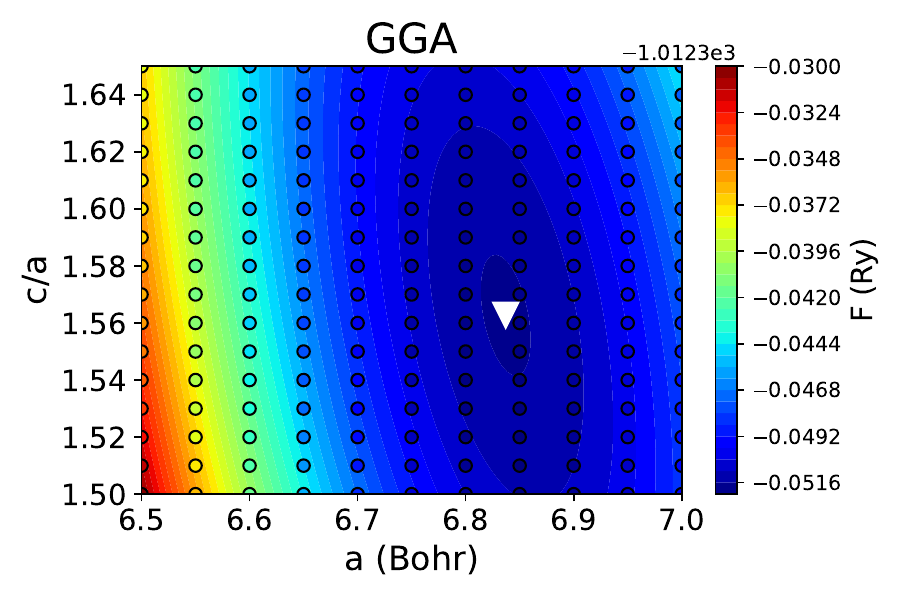} \\     
    \end{tabular}
    \caption{Free energy contour plot of tetragonal FeSe crystal as a function of lattice parameters obtained by GGA+$U$ with U=1, vdW, and GGA. The white triangle shows the minimum of the fitted polynomial function. Similar plots using GGA+$U$ with $U=$4 and 6 are reported in the Supplementary Materials \cite{Suppl}.}
    \label{fig:map}
\end{figure*}

Using finite-temperature density functional theory (DFT) with non-local long-range Van der Waals (vdW) interactions and Hubbard $U$ corrections, we find that (I) the bilayer FeSe, in the absence of external excitations, adopts an antiferromagnetic ground state with a staggered spin configuration, maintaining a net site magnetization of $|M|\sim 2.75~\mu_B$/Fe; (II) this antiferromagnetic order collapses into a paramagnetic state upon increasing the electronic temperature; and (III) this transition is accompanied by an abrupt expansion of the interlayer separation.  
 
Finite-temperature (vdW-corrected) DFT calculations \cite{Hohenberg,Kohn,Anisimov,Liechtenstein,Dudarev,Cococcioni} were performed using scalar-relativistic Ultrasoft Pseudopotentials \cite{UPF}  and the PBEsol \cite{PBEsol} (vdW-DF2 \cite{vdw2}) exchange-correlation functional \cite{PBEsol} as implemented in the Quantum Espresso package v7.2 \cite{QE,QE2}. An effective Hubbard $U$ correction was applied to the Fe $3d$ orbital to account for its correlation effects. The plane-wave cutoff energy for wavefunctions was set to 100 Ry, with an augmentation charge cutoff at 800 Ry. Brillouin zone sampling was performed with a $16 \times 16 \times 12$ $k$-mesh for bulk FeSe, and a $16 \times 16 \times 1$ $k$-mesh for bilayer FeSe. The electronic temperature $T_{el}$ was controlled via the Fermi-Dirac distribution function. 

\textit{Properties of bulk FeSe under ultrafast laser radiation.} 
Bulk FeSe crystallizes in a tetragonal structure with a space group $P4/nmm$ \cite{Wang2020}. To determine the lattice parameters $a$ and $c$ of tetragonal FeSe, we calculated the free energy $F(a, c/a)$ using GGA, vdW and GGA +$U$ (with $U = 1, 4,$ and $6$ eV). For each calculation, all internal atomic positions are fully relaxed (see Fig.~\ref{fig:map}). The resulting free energy values $F$ are then fitted to a polynomial of the form: $F(a, c/a) = m_0 + m_1*a + m_2*(c/a) + m_3*a^2 + m_4*(c/a)^2 + m_5*a*(c/a)$, which allows us to identify the minimum of $F(a, c/a)$ and obtain the corresponding lattice parameters, compared with the experimental values \cite{Koch2019}. Each point in the free-energy contour plots in Fig.~\ref{fig:map} represents a separate DFT calculation with relaxed atomic coordinates. Our calculated parameters are compared with the experimental values in the Supplementary Material \cite{Suppl}.

We studied the behavior of the tetragonal lattice parameters as a function of the electronic temperature by calculating the free energy at different $T_{el} = 0.0136, 0.68, 1.02$ and $1.36$ eV \cite{Suppl}. The same as above, the free energy at different $T_{el}$ was calculated on a mesh of $(a,c/a)$ data points where the atomic coordinates were relaxed at each mesh point. More than a thousand DFT simulations were carried out. We calculated the minimized lattice parameters as described above. We found that both the lattice parameters of $a$ and $c$ and, consequently, the equilibrium volume of the system are enlarged with increasing $T_{el}$. However, $c/a$ decreases with increasing $T_{el}$, indicating that the $xy$-plane lattice parameter $a$ expands faster than $c$ for $T_{el} \leq 1$ eV. Increasing the lattice parameters by enlarging the density of photoexcited electrons is driven by the thermal pressure of the hot electrons, while the lattice is cold, which can be observed from increasing the pressure by $T_{el}$ during an isochoric process \cite{Suppl}.  We also calculated the phonon density of states (DOS) and the projected electronic density of states as a function of $T_{el}$, while the volume is fixed at the equilibrium volume of the system at $T_{el}$. The results of the dynamic lattice calculations denote the softening of the phonon spectra of tetragonal FeSe by increasing $T_{el}$. We previously studied simple metals such as gold and iron and found that photoexcitation of $d$-band electrons causes phonon hardening, which was also reported by experiment and theory \cite{Descamps24, Recoules06}. Although in both cases of FeSe and gold, $d$-band electrons near the Fermi level are photoexcited, the reason that bulk FeSe shows phonon softening with increasing $T_{el}$ but gold phonon hardening could be due to the absence of optical modes in gold as it is a simple metal with a single atom per unit cell. Photoexcitation of $d$-band electrons increases the acoustic mode energies that vanish when $\bf{k}=0$. We observed a softening of the FeSe optical modes with increasing $T_{el}$, which may cause instability of the system and a solid-solid phase transition. Note that the lattice response of the metals upon strong optical excitation is fundamentally different.   

The FeSe crystal shows metallic behavior, as shown by our DFT electronic band structure of the system \cite{Suppl}. The calculated projected electronic density of states (pDOS) indicates that the majority of the DOS vicinity in the Fermi energy belongs to the $d$-orbital of the Fe atom. We calculated the band structure and DOS using vdW, GGA, and GGA+$U$ with $U=1$, and found that all provide similar results for the electronic structure of FeSe with tetragonal structure. We also performed spin-polarized calculations with vdW, and found that the system does not have any finite magnetic moment, which agrees with the experiment and previous calculations. The electronic DOS near the Fermi energy increases by $T_{el}$. The chemical potential of the system increases with $T_{el}$ and the conduction bands shift to lower energies causing a rise in the DOS vicinity of the Fermi energy and decreasing the band-width of $d$-orbital of Fe atom \cite{Suppl}. 

\begin{figure}[htbp!]
    \centering
    \includegraphics[clip, width=0.95\linewidth]{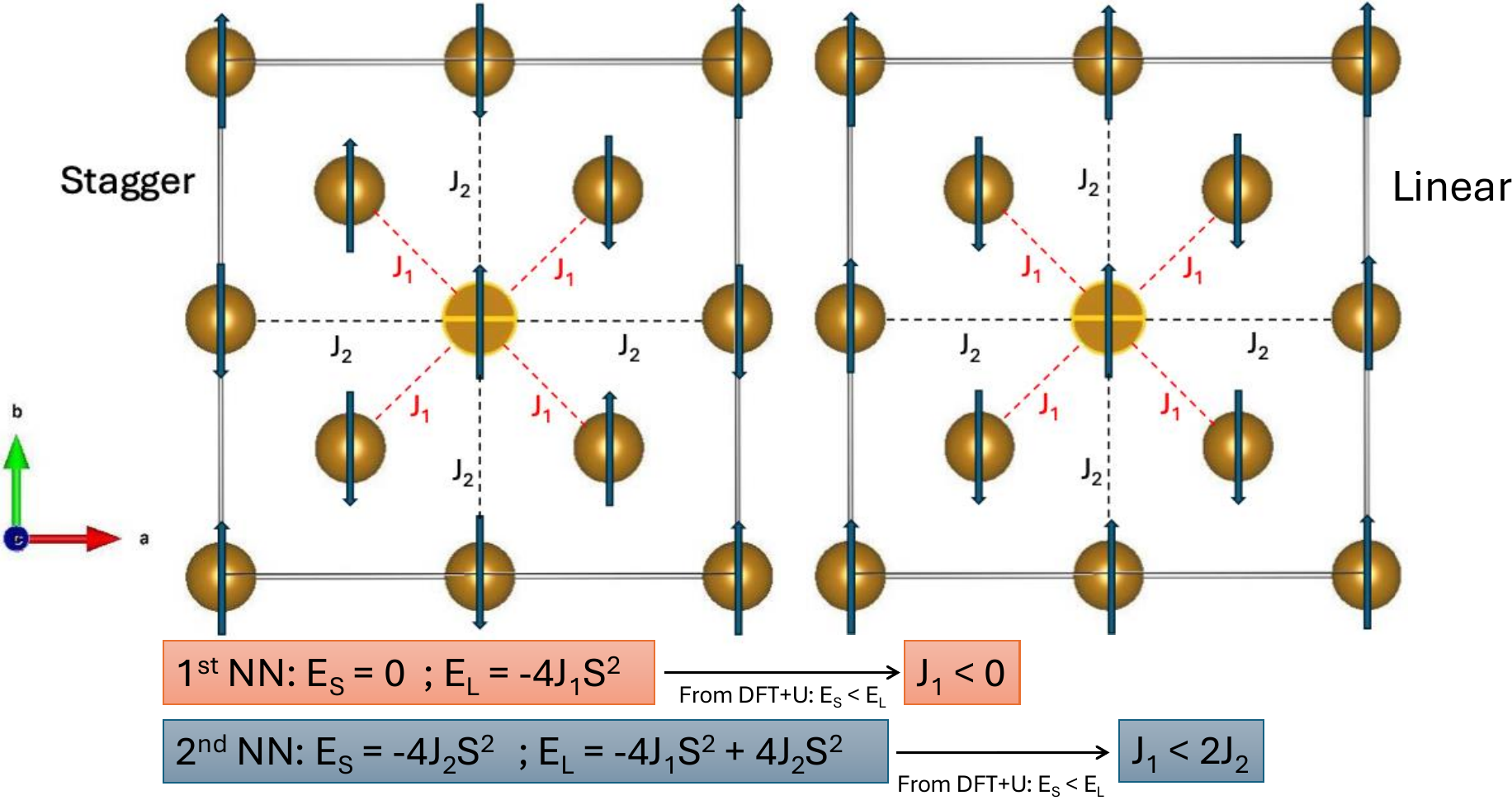}
    \caption{The stagger and linear spin configurations of Fe atoms in each layer of double-layer FeSe. The spin-spin first  (1$^\text{st}$) and second (2$^\text{nd}$) nearest-neighbour (NN) exchange interactions are represented by $\text{J}_1$, and $\text{J}_2$, respectively. E$_\text{S}$ and E$_\text{L}$ are ground state Ising energy of stagger and linear spin configurations, respectively.}
    \label{fig:ising}
\end{figure}

\textit{Emerging anti-ferromagnetic phase}. For the rest of this paper, we present GGA+$U$ results which are obtained with $U=1$. The existence of magnetic ordering for a metallic system substantially suppresses the value of the $U$ parameter \cite{Tesch22}. As we show below, double-layer FeSe adopts metallic antiferromagnetic properties. The structure of the double-layer FeSe and the antiferromagnetic spin configuration (AF) at each layer are illustrated in Supplementary Materials \cite{Suppl}. We considered two antiferromagnetic spin configurations for iron atoms named "linear" and "stagger" (Fig.\ref{fig:ising}). The separation between layers is defined by the parameter $d$. We calculated the Helmholtz $F(d)$ free energy as a function of the distance $d$ between two layers. Simulations were performed for paramagnetic and antiferromagnetic (AF) double-layer FeSe at $T_{el}=0.0136$ eV. A polynomial function of degree three is fitted on the free-energy DFT data to obtain the optimized parameter $d$. 

\begin{figure}[htbp!]
    \centering
    \begin{tabular}{c c}
        \includegraphics[clip,width=0.55\linewidth]{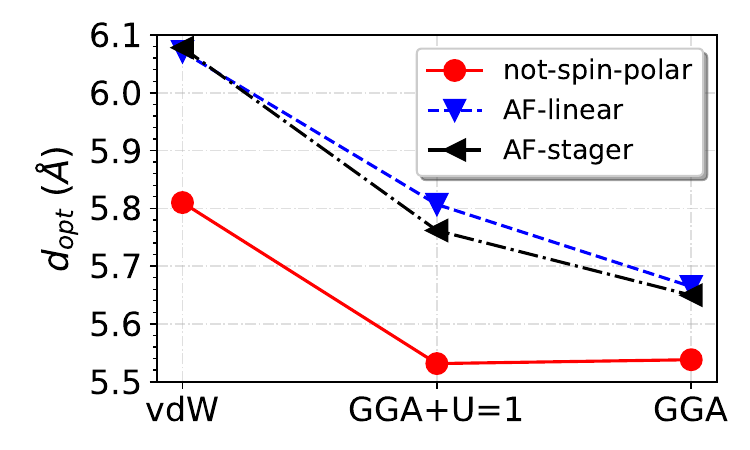} &  
        \includegraphics[clip,width=0.45\linewidth]{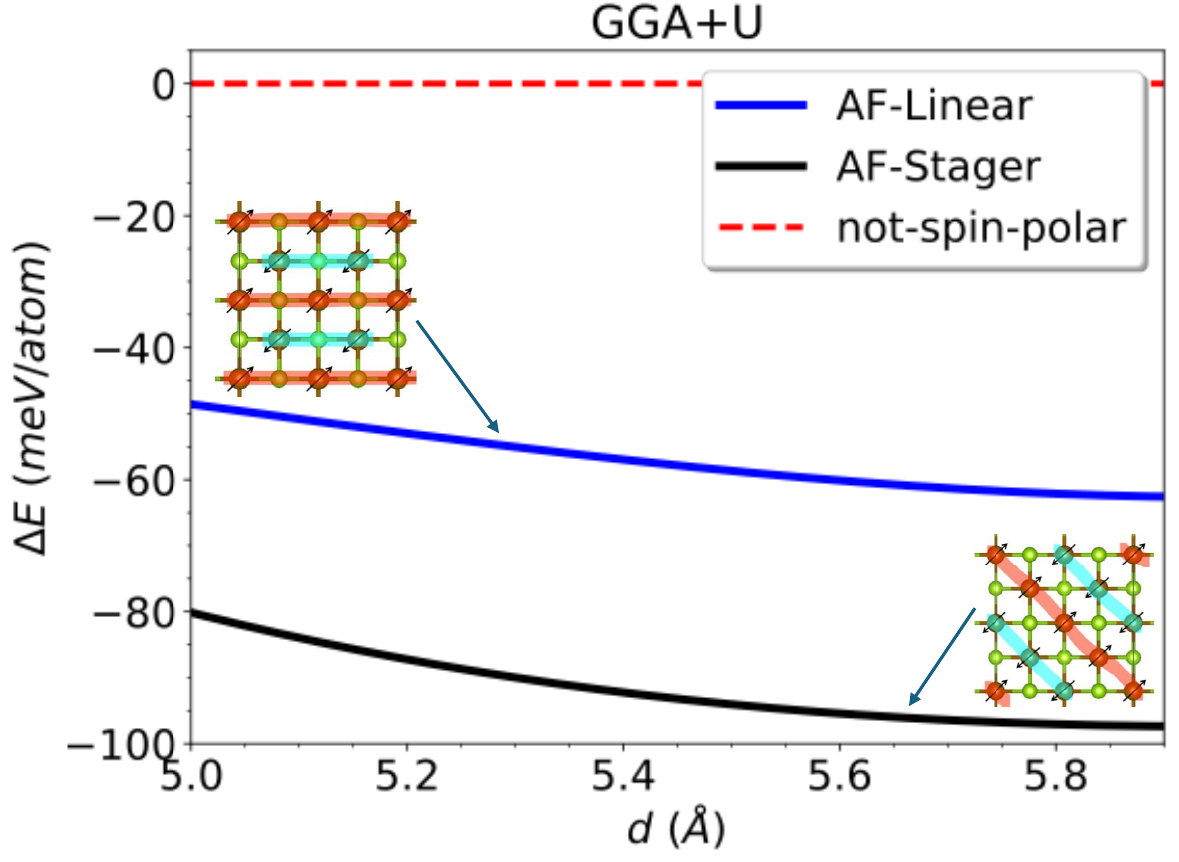} \\ 
        \includegraphics[clip,width=0.5\linewidth]{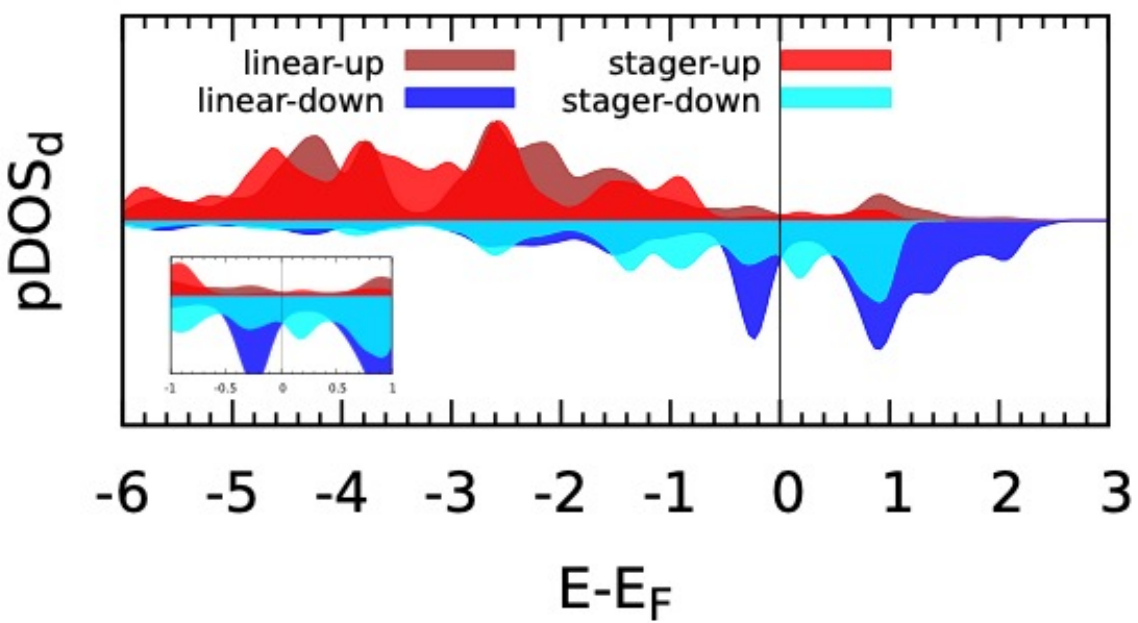}&
        \includegraphics[clip,width=0.45\linewidth]{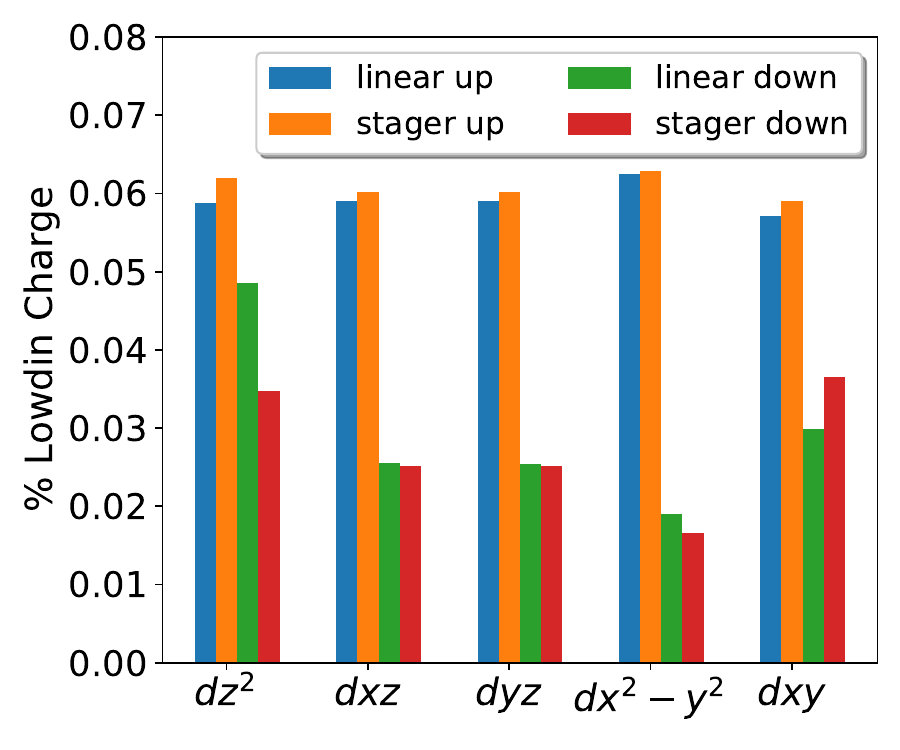}\\
    \end{tabular}
    \caption{(Up left panel) Optimized interlayer separation of double-layer FeSe obtained with vdW, GGA, and GGA$+U=1$ methods. Three systems of not-spin-polarised, antiferromagnetism with linear spin configuration, and antiferromagnetism with stagger spin configuration are considered. The appearance of magnetic ordering increases the distance between two layers. (Upright panel) The energy difference between the studied phases as a function of interlayer separation. The phase diagram is obtained using GGA+$U$ with $U=1$. The antiferromagnetic phase with stagger spin configuration has the lowest energy. The spin-up and spin-down iron atoms are highlighted by red and cyan colours, respectively. (Bottom left panel) The $d$-orbital of spin-up-Fe atom projected density of states (pDOS) for liner and stagger antiferromagnetic configurations. The inset shows the pDOS near the Fermi energy. (Bottom right panel) The percentage of L\"{o}wdin atomic charge of Fe atom located on a layer of double-layer FeSe. GGA+$U$ with $U$=1 are used in both calculations.}
    \label{fig:OptSepar}
\end{figure}

The optimized interlayer separation $d$ obtained by vdW, GGA, and GGA+$U$ are shown in figure~\ref{fig:OptSepar}. It can be observed that the appearance of magnetic ordering increases the distance between two layers of bilayer FeSe owing to the enlarging of the repulsion exchange interaction in the antiferromagnetic system compared with the non-spin polarized case. This conclusion is independent of the exchange-correlation (XC) approximation used in our DFT simulations. The difference between the minimized parameter $d$ predicted by GGA and GGA+$U$ for the non-polarized system is negligible. However, including the $U$ parameter plays an important role in DFT simulations of the antiferromagnetic case, since the GGA+$U$ minimized $d$ parameter is larger than GGA. More importantly GGA+$U$ results show that the linear and stagger spin configurations have a different minimized interlayer separation, whereas the difference between the minimized $d$ parameter of the linear and stagger systems that are obtained by vdW and GGA is negligible. Hence, according to GGA and vdW, the optimized distance between two layers does not depend on the spin configuration of the system, while GGA+$U$ predicts that the stagger spin configuration reduces the separation between layers compared to the linear spin configuration. 

We calculated the free energy difference between antiferromagnetic and non-spin-polarized systems as a function of the distance between layers $d$, as plotted in Fig.~\ref{fig:OptSepar}. The stagger spin configuration has the lowest energy and, therefore, is the most stable phase. We performed similar calculations using vdW and GGA and found the same results\cite{Suppl}. The energy difference between the non-spin-polarized and the antiferromagnetic systems is larger than the free energy difference between linear and stagger antiferromagnetic spin configurations. The energy difference between the stagger antiferromagnetic system and the non-spin-polarized case increases by enlarging the interlayer distance. The phase diagram of bilayer FeSe indicates the emergence of magnetic ordering with a specific stagger spin configuration as a result of the three- to two-dimensional transition. 

The projected density of states (pDOS) of the spin-up Fe atom, calculated by DFT+$U$, for linear and stagger spin configurations, illustrated in Fig.~\ref{fig:OptSepar}, shows the energy difference between the $d$ bands of two spin configurations. Especially in the vicinity of the Fermi energy, the spin-down pDOS of the linear configuration shows a minimum, whereas the spin-down pDOS of the stagger configuration shows a local maximum. The pDOS of the linear spin configuration spreads over a wider energy range with strong localization compared to the stagger spin configuration which is relatively smooth. The difference between the averaged spin-up and spin-down pDOS for the linear configuration is larger than the stagger, suggesting that the averaged Hubbard interaction between spin-up and spin-down electrons in the stagger configuration is weaker than that in the linear configuration.

For better understanding of the difference between linear and stagger spin configurations we analyze the occupation of $d$ orbital energy levels $dz^2$, $dxz$, $dyz$, $dx^2-y^2$, and $dxy$, as shown in the bottom panel of Fig.~\ref{fig:OptSepar}. The plot shows the percentage of L\"{o}wdin atomic charge of Fe atom located on a single layer of double-layer FeSe for each $d$-orbital energy level. The Fe atom has a spin-up majority. The spin-up charge allocation of the $d$ orbital energy levels is almost the same for both antiferromagnetic linear and stagger spin configurations. However, the total spin-down occupation of the $d$ orbital energy levels for the linear configuration is slightly larger than that for the stagger spin configuration. In particular, the $dz^2$ level indicates a clear difference between spin-down L\"{o}wdin localized charge for linear and stagger spin configurations. Our results also show that the absolute magnetization per Fe atom for linear and stagger spin configurations are 2.5, and 2.75 mag Bohr, respectively. Hence, the antiferromagnetic bilayer FeSe tends to spin configuration, which maximizes absolute magnetization. 

We used the Ising Hamiltonian with the first and second nearest-neighbor (NN) interaction to describe the exchange interaction between the spins of Fe atoms at each layer: $H =  \sum_{<ij>} \text{J}_{ij} \text{S}_i \text{S}_j$, where $\text{J}_{ij}=\text{J}_1$, and $\text{J}_2$ for the first (1$^\text{st}$) and second (2$^\text{nd}$) nearest-neighbor exchange interactions, respectively. If we only consider the 1$^\text{st}$NN interaction, the Ising energies of the stagger and linear configuration are E$_\text{S} = 0$ and E$_\text{L}=-4\text{J}_1\text{S}^2$, respectively (Fig.\ref{fig:ising}). The GGA+$U$ phase diagram (Fig.\ref{fig:OptSepar}) shows that E$_\text{S}<$E$_\text{L}$, hence J$_1<0$. The zero energy of the 1$^\text{st}$NN interaction explains the larger local magnetic moment in the stagger configuration.  Taking into account the 1$^\text{st}$NN and the 2$^\text{nd}$NN interactions gives E$_\text{S}=-4\text{J}_2\text{S}^2$ and E$_\text{L}=-4\text{J}_1\text{S}^2+4\text{J}_2\text{S}^2$. Since E$_\text{S}<$E$_\text{L}$ as predicted by GGA+$U$, one can easily find that J$_2>$J$_1/2$. 

\begin{figure}[!htpb]
    \centering
    \begin{tabular}{cc}
       \includegraphics[width=0.45\linewidth]{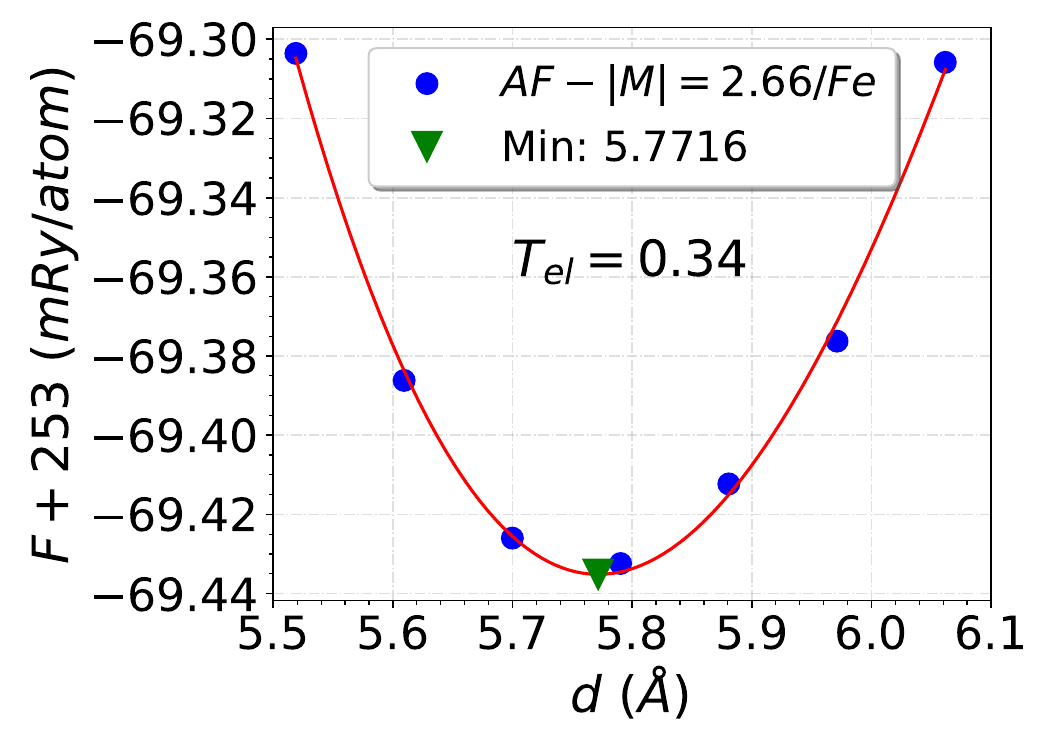}  & 
       \includegraphics[width=0.45\linewidth]{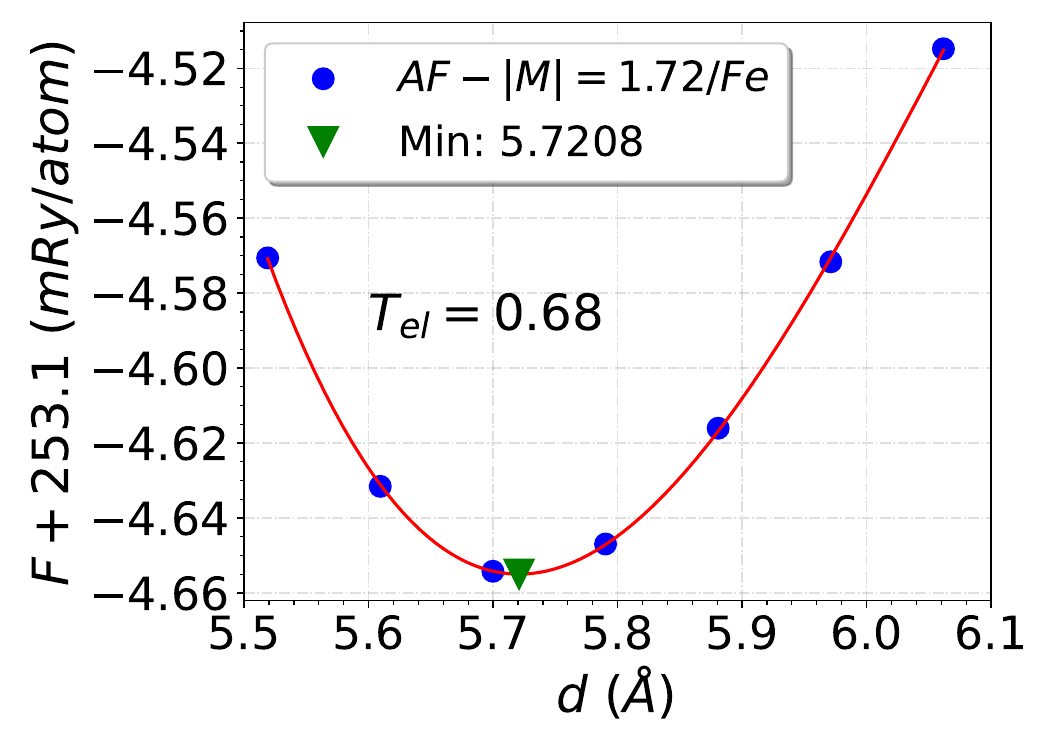}  \\
       \includegraphics[width=0.45\linewidth]{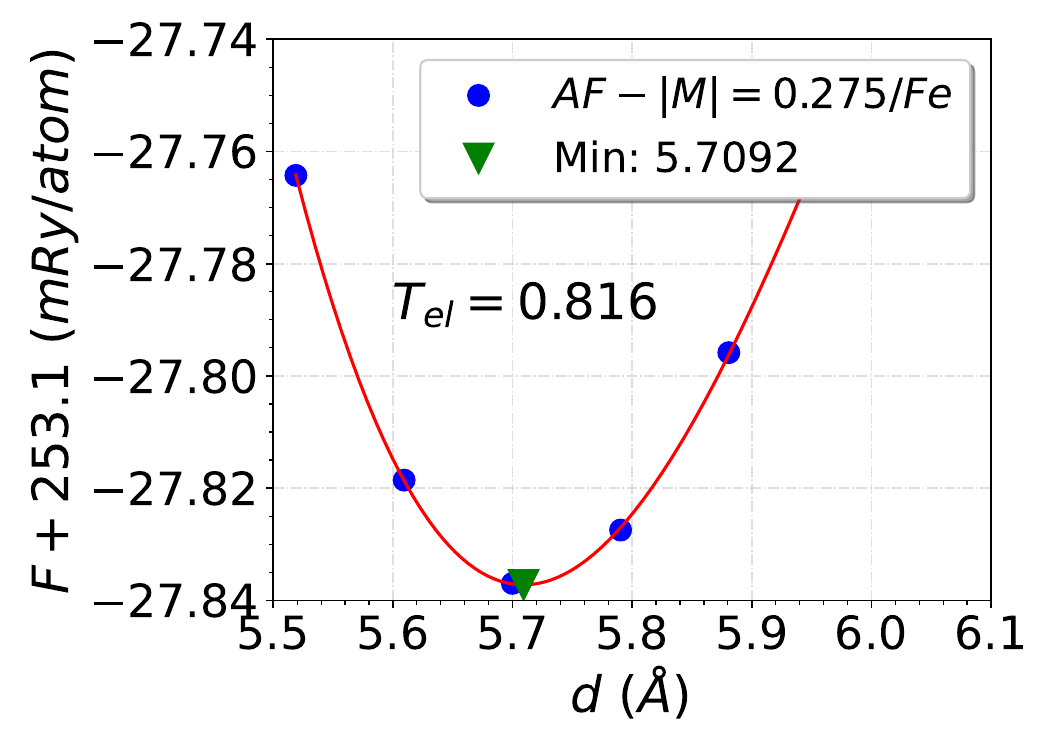}  &
       \includegraphics[width=0.45\linewidth]{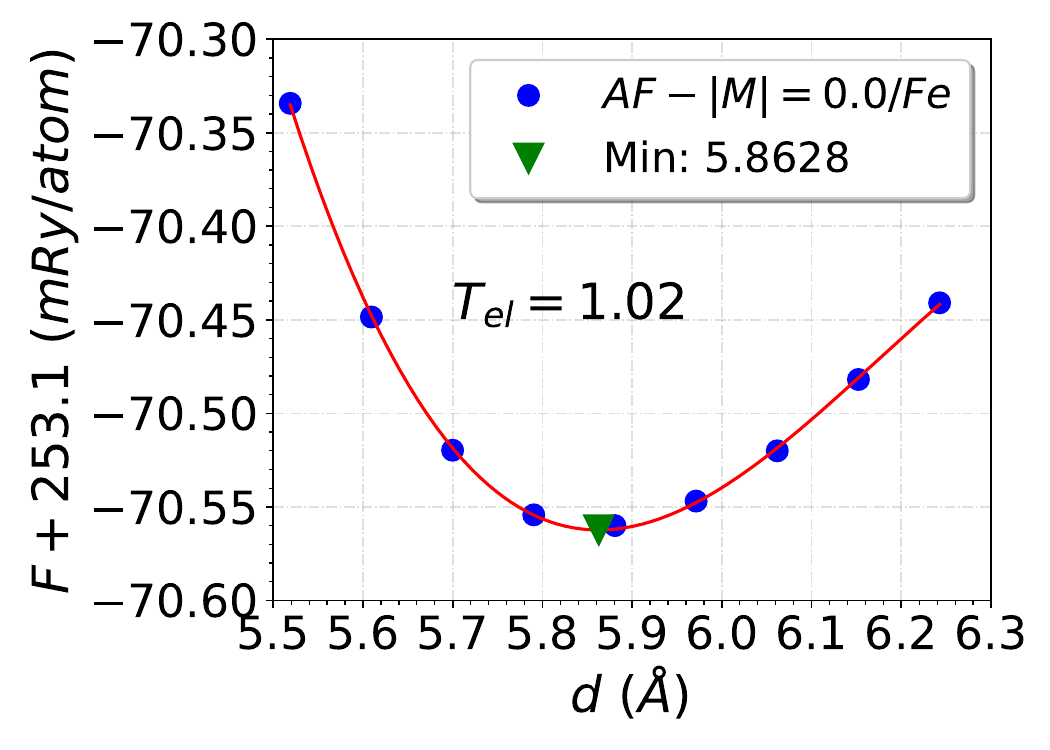}  \\
       \includegraphics[width=0.45\linewidth]{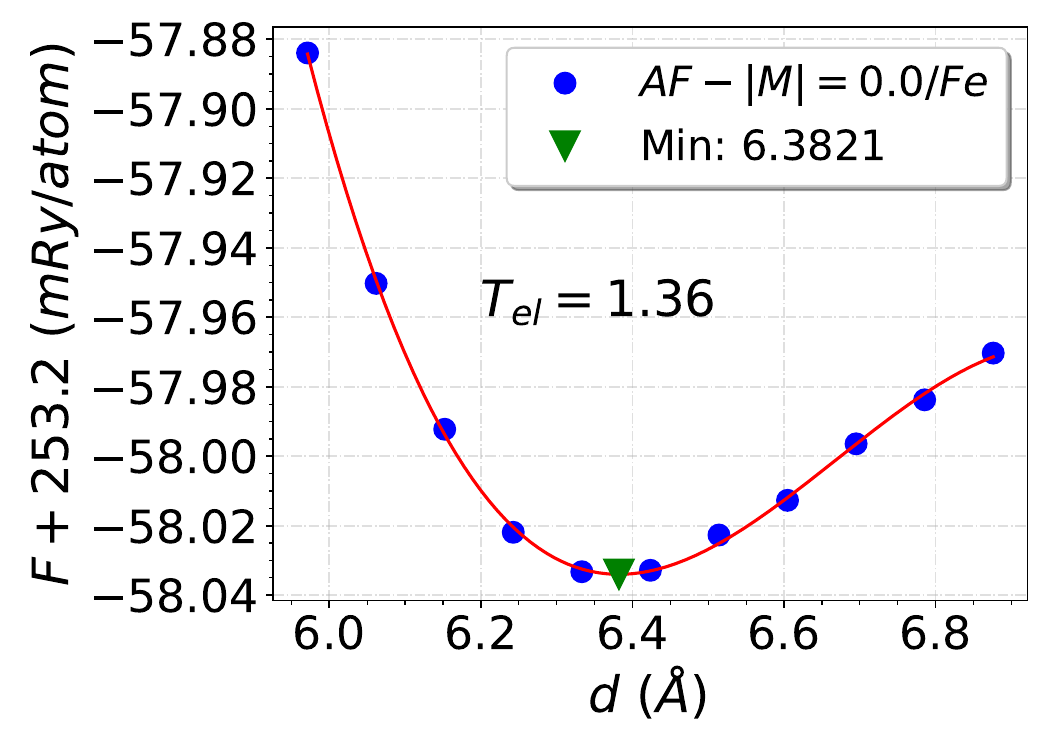}  &
       \includegraphics[width=0.45\linewidth]{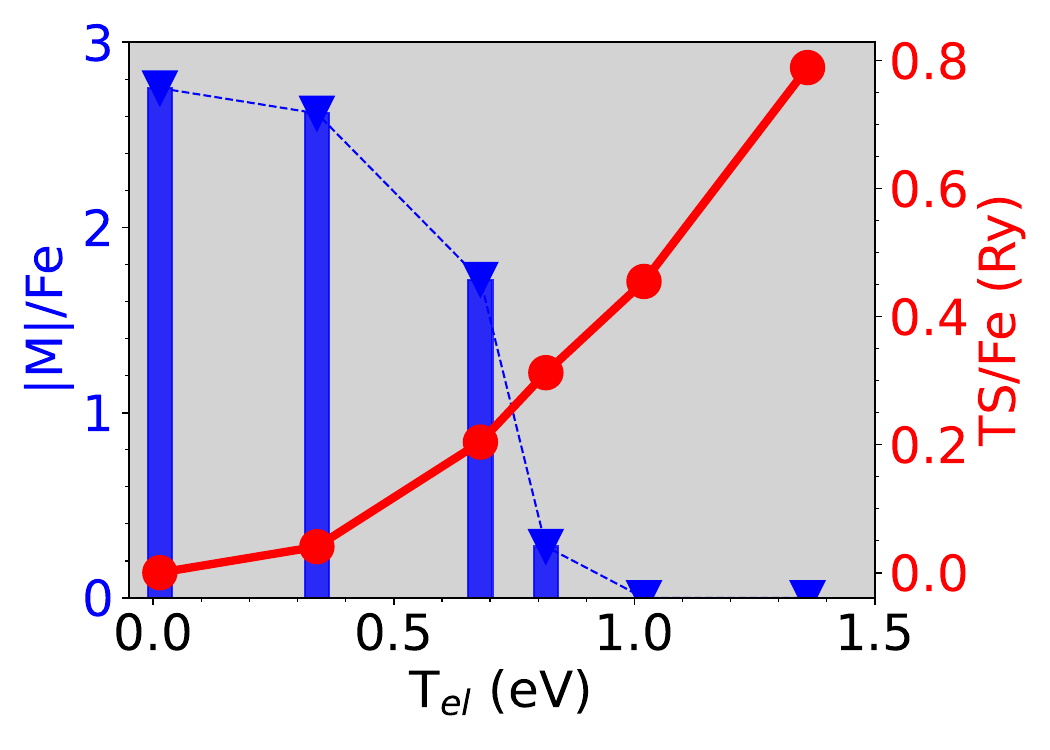}  \\
    \end{tabular}
    \caption{The Helmholtz free energy $F$ as a function of interlayer separation $d$ of anti-ferromagnetic double-layer FeSe with stagger spin configuration. The system is fully optimized at electronic temperatures $T_{el} = 0.34, 0.68, 0.816, 1.02, 1.36 ~eV$. The minimized $d$ is obtained by fitting a polynomial on the data points which are calculated using GGA+$U$ with $U=1$. The last plot on the right side shows the absolute magnetization $|M|$ and the electronic entropy $TS$ per iron atom as a function of the electronic temperature T$_{el}$. }
    \label{fig:collapse}
\end{figure}

\textit{Magnetic collapse and interlayer separation expansion}. 
In this section, we consider the stagger spin configuration, since it has the lowest thermodynamic energy. We investigated the effects of photo-excited electrons on the magnetic, electronic and structural properties of the metallic antiferromagnetic double layer FeSe by calculating the Helmholtz free energy as a function of electronic temperature $T_{el}$. The atomic coordinates of the system at each electronic temperature and interlayer separation were fully relaxed. Figure~\ref{fig:collapse} shows the Helmholtz free energy as a function of interlayer separation that was calculated at $T_{el} = 0.34, 0.68, 0.816, 1.02$ and $1.36$ eV. 

\begin{figure}
    \centering
    \includegraphics[width=0.9\linewidth]{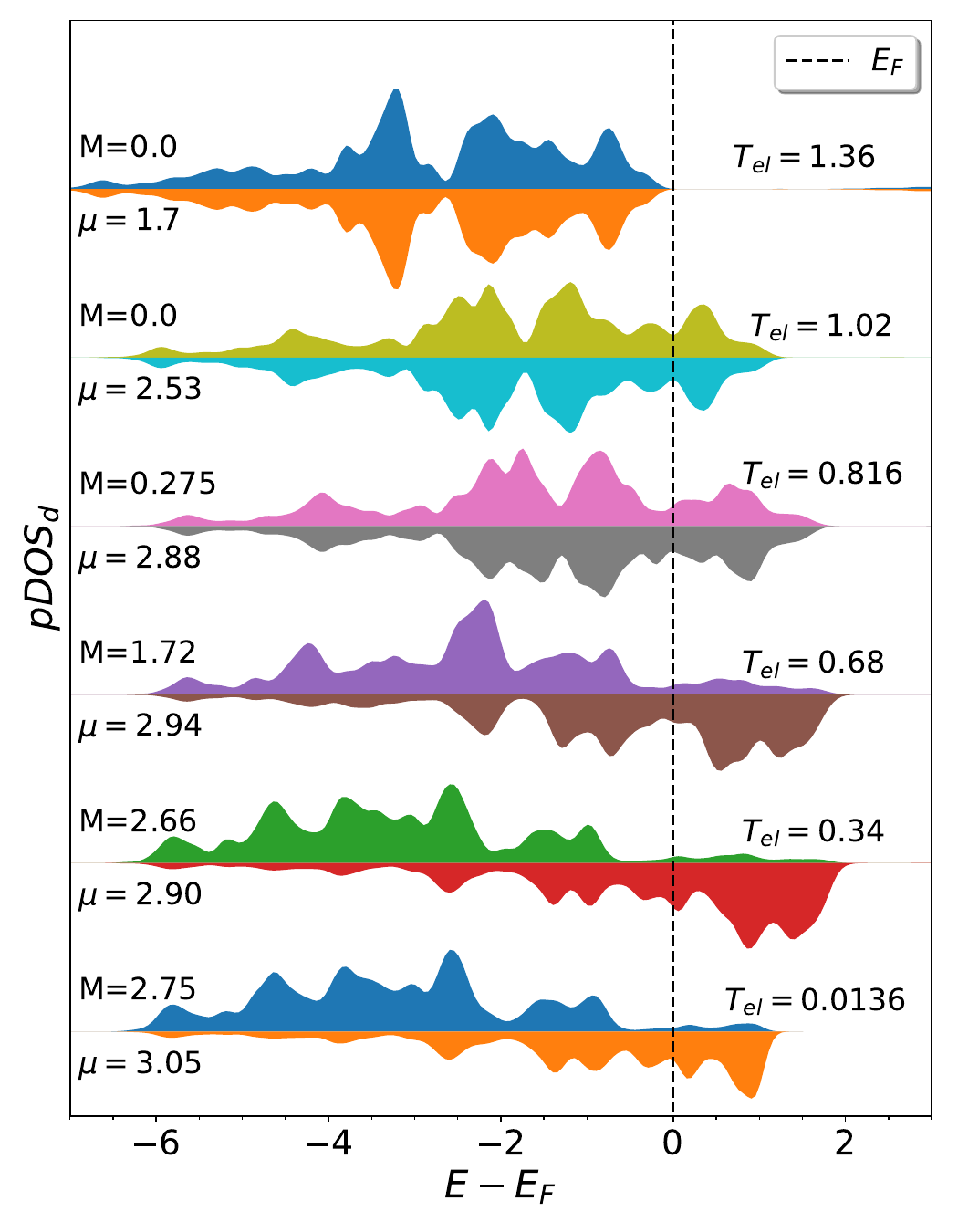}
    \caption{Project density of states of d-orbital ($pDOS_d$) for an iron atom with spin up majority that is located on a layer of double-layer FeSe with stagger spin configuration. The electronic temperature $T_{el}$ in eV, the chemical potential $\mu$ in eV, and absolute local magnetization M, are shown for each pDOS$_d$. The Fermi energy $E_F$ is defined as the chemical potential at zero temperature.}
    \label{fig:pdosvsT}
\end{figure}

We found that demagnetization, which is introduced by increasing the density of photoexcited electrons and the entropy of hot electrons, occurs within the narrow energy window of $0.82-1.0~eV$ (Fig.~\ref{fig:collapse}). The decrease in absolute local magnetic moment $|M|$ of the iron atom shows two drops at electronic temperatures of $\sim$ 0.68 and 0.82. In the first drop, the local magnetic moment of iron decreases from 2.66 to 1.7, and in the second drop, which is steeper than the first drop, $|M|/Fe$ reduces from 1.72 to 0.275. The reason for these two drops can be observed from the projected DOS of the iron $d$ orbital (pDOS$_d$) as illustrated in Fig.\ref{fig:pdosvsT}, and the Stoner model. The magnetic properties of antiferromagnet double layer FeSe, which depends on $T_{el}$, can be related to the Stoner excitations from the occupied spin-up $d$-band of iron atom to the unoccupied spin-down $d$-band, which can be observed from increasing the pDOS$_d$ spin-down electrons near the Fermi level (Fig.~\ref{fig:pdosvsT}). These excitations reduce the magnitude of the absolute magnetization, so that finally at a critical electronic temperature $0.82 <T_c< 1.0$ eV the system becomes paramagnetic. Using a small value of $U=1$ in our simulations means that the system was considered weakly correlated, due to its metallic screening effects,  and therefore the only possible magnetic excitations are spin flips. 

Photoexcited electrons have double effects on the interlayer separation. As illustrated in Fig.~\ref{fig:collapse}, by gradually increasing the electronic temperature, the system goes through a ferrmagnetic-paramagentic phase transition, resulting in a decrease in the distance between two layers. The optimized distance between two layers just before demagnetization is the smallest since the system adopts paramagnetic properties. After demagnetization, the interlayer separation expands with the electronic temperature, which is driven by the thermal pressure of the hot electrons. Note that the nonlocalized electron heat capacity is much lower than the lattice heat capacity. Hence, the electrons can be heated to temperatures much higher than the Fermi temperature $T_F$ while the lattice is still cold.

\textit{Conclusion.} We studied the effects of photoexcited hot electrons on the electronic, magnetic, and structure properties of bulk and two-dimensional double layer FeSe. Our first principal phase diagrams were obtained using three different XC approximations which belong to the categories of the DFT functionals GGA, GGA+$U$, and vdW. We concluded that our main results are independent of the XC approximation.  (i) Although bulk FeSe is a paramagnetic metal, bilayer FeSe is an antiferromagnetic metal with a unique stagger spin configuration which maximizes the local magnetic moment of iron atom on each layer; (ii) phoexcitation of the electrons near the Fermi energy and their electronic temperature, which can be controlled by laser radiation, introduce the antiferromagnetic-paramagnetic phase transition in double-layer FeSe. The distance between two layers initially decreases with the electronic temperature during the demagnetization process, but it expands by increasing the electronic temperature of the bilayer paramagnetic FeSe. The interlayer spacing influences the properties of 2D magnetic systems, including FeSe.  From an experimental point of view, it is challenging to control the interlayer distance bidirectionally and reversibly. Since rapid demagnetization occurs in a narrow energy window of $\sim 0.2$ eV, by precisely timing the laser pulses, it is possible to coherently control the electronic and magnetic properties of 2D-FeSe for applications in ultrafast switches and information processing.

We acknowledge the support of the Leverhulme Trust under the grant agreement RPG-2023-253.

\bibliography{mainbib}

\end{document}